\documentclass[
fleqn,pulished]{elsart}
\usepackage{amsmath,amssymb}

\def\be{\begin{equation}}
\def\ee{\end{equation}}
\def\bea{\begin{eqnarray}}
\def\eea{\end{eqnarray}}
\def\nn{\nonumber}

\begin{document}
\begin{flushright}
hep-th/0603082 
\end{flushright}

\begin{frontmatter}

\title{\bf Hawking radiation via tunnelling from higher dimensional
Reissner-Nordstr\"{o}m-de Sitter black holes}

\author[CPST]{Shuang-Qing Wu\corauthref{cor}}\ead{sqwu@phy.ccnu.edu.cn},
\corauth[cor]{Corresponding author.}
\author[IoPP]{Qing-Quan Jiang}\ead{jiangqingqua@126.com}

\address[CPST]{College of Physical Science and Technology, Central China Normal University,
Wuhan, Hubei 430079, People's Republic of China}
\address[IoPP]{Institute of Particle Physics, Central China Normal University,
Wuhan, Hubei 430079, People's Republic of China}

\begin{abstract}
Recent work that treats the Hawking radiation as a semi-classical tunnelling process from the
four-dimensional Schwarzschild and Reissner-Nordstr\"{o}m black holes is extended to the case
of higher dimensional Reissner-Nordstr\"{o}m-de Sitter black holes. The result shows that the
tunnelling rate is related to the change of Bekenstein-Hawking entropy and the exact radiant
spectrum is no longer precisely thermal after considering the black hole background as dynamical
and incorporating the self-gravitation effect of the emitted particles when the energy conservation
and electric charge conservation are taken into account.
\end{abstract}

\end{frontmatter}

\newpage

\section{Introduction}

Hawking's remarkable discovery \cite{SWH} in 1974 that black holes can radiate thermally reconciled
a serious contradiction among General Relativity, Quantum Mechanics and Thermodynamics at that time
and put the first law of black hole thermodynamics on a solid fundament. At a big cost, however,
this discovery also caused another controversial problem: what happen to information during the
black hole evaporation? This notable problem is now often called as the information loss paradox
\cite{ILP}, since the radiation with a pure thermal spectrum has no way to be recovered after black
holes have disappeared completely. Initially Steven Hawking and Kip Thorne thought that information
is lost during the process of black hole evaporation, but John Preskill believed that information
is not lost and can get out of the black hole. In 2004, Steven Hawking \cite{ILP} changed his old
opinion and argued that the information could come out if the outgoing radiation were not exactly
thermal but had subtle corrections.

Recently Parikh and Wilzcek \cite{PW} presented a greatly simplified model (based upon the previous
developments by Per Kraus, \textit{et al}. \cite{KKW}) to implement the Hawking radiation as a
semi-classical tunnelling process from the event horizon of the four-dimensional Schwarzschild and
Reissner-Nordstr\"{o}m black holes by treating the background geometries as dynamical and incorporating
the self-gravitation correction of the radiation. The radiant spectra that they derived under the
consideration of energy conservation give a leading-order correction to the emission rate arising
from the loss of mass of black holes, which corresponds to the energy carried by the radiated quanta.
Their result shows that the actual emission spectrum of black hole radiation deviates from strictly
pure thermality, which might serve as a potential mechanism to resolve the information loss paradox.
Following this program, a lot of recent work (see Ref. \cite{JW} and references therein) has been
focused on extending this semi-classical tunnelling method to various cases of black holes such as
those in de Sitter \cite{dSTR,AM,RNdSTR}, anti-de Sitter \cite{JW,HK,AdSTR} space-times, charged
\cite{RNdSTR,ZZ} and rotating \cite{RBHTR} black holes as well as other cases \cite{ECV}. But
relatively less research \cite{AM,HK,RZG} in higher dimensions is touched upon.

In this Letter, we shall investigate the Hawking radiation via tunnelling from higher dimensional
Reissner-Nordstr\"{o}m black holes in de Sitter space-time. The subject is mainly motivated by the
fact that recent brane-world scenarios \cite{BW} predict the emergence of a TeV-scale gravity in the
higher-dimensional theories, thus open the possibility to test Hawking effect and to explore extra
dimensions by making tiny black holes in the high-energy colliders \cite{LHC} (LHC) to be running
in the next year. Within this context, recently there is much attention \cite{HDBHR} being paid to
examine the various properties, especially Hawking radiation of the higher-dimensional black holes. Our
interest on this topic is also due to the following two important aspects: (1) In the light of recent
astronomical observation data for supernova of type Ia and the power spectrum of CMB fluctuations,
it has been suggested that our universe is in a phase of accelerating expansion \cite{CAE} and will
asymptotically approach a de Sitter space. That is to say, the cosmological constant in our present
universe might be a small but positive one. This realization has sparked a sense of urgency in resolving
the quantum-gravitational mysteries of de Sitter space. (2) A conjectured de Sitter/conformal field
theory (CFT) correspondence \cite{dSCFT} defined in a manner analogous to the very successful AdS/CFT
correspondence \cite{AdSCFT} has been proposed recently that there is a dual between quantum gravity
on a de Sitter space and a Euclidean conformal field on a boundary of the de Sitter space. Much work
has been done on finding a holographic description of de Sitter space and establishing an analogous
duality. Although there has been considerable success along this line, the proposed duality is still
marred by various ambiguities. Hence as an intermediate step, it may be appropriate to reinforce our
understanding of de Sitter space at a semi-classical level by considering the de Sitter radiation as
a semi-classical tunnelling process.

On the grounds of all described above, it becomes obvious that the further study of black holes in
higher dimensions is of great important. Of particular interest is the case of Hawking radiation of
the higher dimensional black holes. The aim of this Letter is to present a reasonable extension of
the Kraus-Parikh-Wilczek's semi-classical tunnelling framework in the four-dimensional spherically
symmetric space-times to the case of higher dimensional Reissner-Nordstr\"{o}m black holes in de
Sitter space-time. Moreover, we shall investigate the tunnelling radiation of the massless uncharged
particles and massive charged particles across the black hole event horizon and the cosmological
horizon, respectively. In addition, their possible phenomenological implications to the brane-world
emission rate is briefly outlined.

Our Letter is organized as follows. After introducing the metric of higher dimensional
Reissner-Nordstr\"{o}m-de Sitter black holes and their Painlev\'{e}-type extensions in Section 2, we
discuss Hawking radiation of massless uncharged particles and massive charged particles as a semi-classical
tunnelling process from the black hole event horizon and the cosmological horizon in Sections 3 and 4,
respectively, where the tunnelling rate and emission spectrum are explicitly computed in each case.
The Letter is ended up with a brief remark.

\section{Higher dimensional Reissner-Nordstr\"{o}m-de Sitter black holes and their Painlev\'{e}-type
extensions}

The line element of ($n + 2$)-dimensional Reissner-Nordstr\"{o}m-de Sitter black holes with a positive
cosmological constant $\Lambda = n(n+1)/(2l^2)$ and the electro-magnetic one-form potential are given by
\cite{CRG,HDBHS}
\bea
ds^2 &=& -f dt_R^2 +f^{-1}dr^2 +r^2d\Omega_n^2 \, , \label{HDRNdS} \\
\mathcal{A} &=& \pm \frac{Q}{(n-1)V_nr^{n-1}}dt \, ,
\label{EMP}
\eea
where
\be
f = f(M, Q, r) = 1 -\frac{\omega_n M}{r^{n-1}}
+\frac{\omega_n Q^2}{2(n-1)V_nr^{2n-2}} -\frac{r^2}{l^2} \, ,
\quad~ \omega_n = \frac{16\pi}{nV_n} \, .
\ee
Here $M$ and $Q$ are the mass and electric charge of the black hole, respectively. $l$ is the curvature
radius of de Sitter space, $V_n$ denotes the volume of a unit $n$-sphere $d\Omega_n^2$. (Units $G_{n+2}
= c = \hbar = 1$ are adopted throughout this article.)

When $M = Q = 0$, the solution (\ref{HDRNdS}) reduces to the pure de Sitter space with a cosmological
horizon $r = l$ which may be very large according to the existing knowledge of the cosmological constant.
When the positive mass $M$ increases with the charge $Q$ decreased or fixed, a black hole horizon occurs
and increases its size, while the cosmological horizon shrinks. For a special value of $M$ and $Q$, the
black hole horizon will touch the cosmological horizon. Besides, when the cosmological constant vanishes,
the black hole has an outer/inner horizon located at
\be
r_{\pm}^{n-1} = \frac{\omega_n}{2}\Big[M \pm\sqrt{M^2 -\frac{nQ^2}{8\pi (n-1)}}\Big] \, .
\ee

For the general case in higher dimensions ($n\geq 2$), the horizons are determined by the equation
$f(M, Q, r_h) = 0$ which is of ($2n$)-order so it in general has ($2n$)-roots. Typically for $n\geq 2$,
there will be three positive (real) roots of $f(M, Q, r_h)$, with the outermost root describing a
cosmological horizon, and the remaining pair describing inner and outer black-hole horizons. Just
as did in the four-dimensional case \cite{RNdSTR}, here we are only interested in two distinct real
roots of them: the largest one is the cosmological horizon (CH) $r_c$, the smaller one corresponds
to the event horizon (EH) $r_+$ of the black hole. The explicit forms of these solutions are not
needed for our discussions made here and are not illuminating though a detailed analysis \cite{dSCV}
may be of some interest.

Obviously, a key trick to apply the semi-classical tunnelling analysis is to introduce a coordinate
system that is well-behaved at the horizons. Thus to remove the coordinate singularities at the
horizons, one must perform a Painlev\'{e}-type coordinate transformation in arbitrary dimensions
\be
dt_R = dt \mp \frac{\sqrt{1 -f}}{f}dr \, ,
\ee
under which the electric potential (\ref{EMP}) retains its previous form modulo a gauge transformation,
and the line element (\ref{HDRNdS}) is transformed to the following Painlev\'{e}-like form
\be
ds^2 = -f dt^2 \pm 2\sqrt{1 -f}dtdr +dr^2 +r^2d\Omega_n^2 \, ,
\label{PLE}
\ee
where the plus (minus) sign denotes the space-time line element of the outgoing (ingoing) particles
across the EH and the CH, respectively. The new metric (\ref{PLE}) is very advantageous for us to
investigate the radiation of particles tunnelling across the horizons and to do an explicit computation
of the tunnelling probability.

First, let us work with the metric in the new form (\ref{PLE}) and obtain the radial geodesics of the
massless uncharged particles that follow the radial null geodesics, which is given by
\be
\dot{r} = \frac{dr}{dt} = \pm 1 \mp \sqrt{1 -f} \approx \pm\kappa_h\big(r -r_h\big) \, , \quad~
\kappa_h = \kappa_h(M, Q, r_h) = \frac{1}{2}\frac{\partial f}{\partial r}\Big|_{r = r_h} \, .
\label{RNG}
\ee
where the upper (lower) sign denotes the radial null geodesics of the outgoing (incoming) radiation
from the EH (CH), namely, $r_h = r_+$, $r_c$, respectively.

Next, it is known that different from the radial null geodesics of the massless uncharged particles, the
trajectory followed by the charged massive particles is not light-like, it does not follow the radially
light-like geodesics when it tunnels across the horizon. By treating the charged massive particle as a
sort of de Broglie s-wave, its trajectory can be approximately determined as \cite{RNdSTR,ZZ}
\be
\dot{r} = \frac{dr}{dt} = -\frac{g_{tt}}{2g_{tr}}
 = \pm \frac{f}{2\sqrt{1 -f}} \approx \pm \kappa_h\big(r -r_h\big) \, ,
\label{MCRGL}
\ee
where the plus (minus) sign represents the radial geodesics of the charged particles tunnelling across
the EH (CH), respectively.

In the following, we shall investigate Hawking radiation via tunnelling across the EH and CH, respectively,
and calculate the tunnelling rate from each horizon. The overall picture of tunnelling radiation for the
metric is very complicated, because radiation is both propagating inwards from the CH and outwards from
the EH. A formal, complete analysis must consider both of these effects, and there would undoubtedly be
scattering taking place between the black hole horizon and cosmological horizons \cite{AM}. To simplify
the discussion, we will consider the outgoing radiation from the EH, and ignore the incoming radiation
from the CH when we deal with the black hole event horizon. While dealing with the CH case, we shall only
consider the incoming radiation from the CH and ignore the outgoing radiation from the EH. This assumption
is reasonable as long as the two horizons separate away very large from each other, since the radius of
the cosmological horizon is very large due to a very small cosmological constant, and the black hole event
horizon considered here is relatively very small because the Hawking radiation can take an important effect
only for tiny black hole typical of $1\sim10$-Tev energy in the brane-world scenario. On the other hand,
because Hawking radiation is a kind of quantum effect, it can be neglected and may not be observed for
an astrophysical black hole with typical star mass about $10M_{\odot}$.

\section{Radiation of massless uncharged particles as tunnelling}

Before proceeding to discuss Hawking radiation of charged massive particles as a semi-classical tunnelling
process, let us first consider the uncharged radiation outgoing from the EH. We adopt the picture of a pair
of virtual particles spontaneously created just inside the horizon. The positive energy virtual particle can
tunnel out and materialize as a real particle escaping classically to infinity, its negative energy partner
is absorbed by the black hole, resulting in a decrease in the mass of the black hole, at the same time,
the size of EH will reduce and the radius of CH will enlarge. Since the emitted particle can be treated
as a shell of energy $\omega$, Eqs. (\ref{PLE}) and (\ref{RNG}) should be modified when the particle's
self-gravitation is incorporated. Taking into account the energy conservation only, the mass parameter
in these equations will be replaced with $M\to M -\omega$ when the particle of energy $\omega$ tunnels
out of the EH. So the radial null geodesics of the uncharged particles tunnelling out from the EH is
\be
\dot{r} = 1 -\sqrt{1 -f(M -\omega, Q, r)}
\approx \kappa_+(M -\omega, Q, r_+)\big(r -r_+\big) \, .
\label{RNLEH}
\ee
The imaginary part of the action of the uncharged particles can be expressed as
\be
\mathrm{Im} S = \mathrm{Im}\int_{r_i}^{r_f}\int_0^{P_r}dP_r^{\prime}dr
= \mathrm{Im}\int_{r_i}^{r_f}\int_M^{M-\omega}\frac{dr}{\dot{r}}d(M-\omega^{\prime}) \, ,
\label{IAEH}
\ee
where we have used the Hamilton equation
\be
\dot{r} = \frac{d(M-\omega)}{dP_r} \, ,
\ee
and denoted $r_{i}$ and $r_{f}$ the locations of the EH before and after the particle of energy $\omega$
tunnels out, respectively.

Substituting Eq. (\ref{RNLEH}) into Eq. (\ref{IAEH}) and switching the order of integration, the imaginary
part of the action can be easily evaluated by deforming the contour around the single pole $r = r_+^{\prime}$
at the EH and reads
\bea
\mathrm{Im}S &=& \mathrm{Im}\int_M^{M -\omega}\int_{r_i}^{r_f}
\frac{dr}{1 -\sqrt{1 -f(M -\omega^{\prime}, Q, r)}}d(M -\omega^{\prime}) \nn \\
&=& -\int_M^{M -\omega} \frac{\pi}{\kappa_+(M -\omega^{\prime}, Q, r_+^{\prime})}
d(M -\omega^{\prime}) \nn \\
&=& -\frac{2\pi}{n\omega_n}\big(r_f^n -r_i^n\big) = -\frac{1}{2}\Delta S_{EH} \, ,
\eea
where $S_{EH} = V_nr_+^n/4 = 4\pi r_+^n/(n\omega_n)$ is the Bekenstein-Hawking entropy of the EH.

Since the geometrical optical limit and the ``s-wave'' approximation can be used here, using the
semi-classical WKB method the tunnelling probability is found to be related to the imaginary part
of the action via $\Gamma \sim e^{-2\mathrm{Im}S}$, so the tunnelling rate at the EH is
\be
\Gamma \sim e^{-2\mathrm{Im}S} = e^{\frac{4\pi}{n\omega_n}(r_f^n -r_i^n)} = e^{\Delta S_{EH}} \, ,
\label{TRE}
\ee
where $\Delta S_{EH} = S_{EH}(M -\omega, Q) -S_{EH}(M, Q)$ is the difference of black hole entropy after
and before the particle emission. Obviously, the derived emission spectrum actually deviates from pure
thermality, perfectly generalizing those obtained in Refs. \cite{PW,KKW,AM}.

Consider now the case of the massless uncharged radiation incoming from the CH, Eqs. (\ref{PLE}) and
(\ref{RNG}) should also be modified when the particle's self-gravitation is incorporated. After taking
into account the energy conservation, the mass parameter in  these equations will be replaced with
$M\to M +\omega$ when the particle of energy $\omega$ tunnels into the CH, and the radial null geodesics
of the uncharged particles tunnelling into the CH becomes
\be
\dot{r} = -1 +\sqrt{1 -f(M +\omega, Q, r)}
\approx -\kappa_c(M +\omega, Q, r_c)\big(r -r_c\big) \, .
\ee
Applying the Hamilton equation
\be
\dot{r} = -\frac{d(M+\omega)}{dP_r} \, ,
\ee
the imaginary part of the action of the uncharged particles can be computed in a similar manner by
deforming the contour around the single pole $r = r_c^{\prime}$ at the CH
\bea
\mathrm{Im}S &=& -\mathrm{Im}\int_{r_{ic}}^{r_{fc}} \int_{-M}^{-(M +\omega)}
\frac{dr}{\dot{r}} d(M +\omega^{\prime})\nn \\
&=& \mathrm{Im}\int_{-M}^{-(M +\omega)}\int_{r_{ic}}^{r_{fc}}
\frac{dr}{1- \sqrt{1 -f(M +\omega^{\prime}, Q, r)}}d(M +\omega^{\prime}) \nn \\
&=& -\int_{-M}^{-(M +\omega)} \frac{\pi}{\kappa_c(M +\omega^{\prime}, Q, r_c^{\prime})}
d(M +\omega^{\prime}) \nn \\
&=& -\frac{2\pi}{n\omega_n}\big(r_{fc}^n -r_{ic}^n\big) \, ,
\eea
where $r_{ic}$ and $r_{fc}$ are the locations of the CH before and after the uncharged particle tunnels
into the CH, respectively. In terms of the Bekenstein-Hawking entropy of the CH: $S_{CH} = V_nr_c^n/4 =
4\pi r_c^n/(n\omega_n)$, the tunnelling rate at the CH is
\be
\Gamma \sim e^{-2\mathrm{Im}S} = e^{\frac{4\pi}{n\omega_n}(r_{fc}^n -r_{ic}^n)} = e^{\Delta S_{CH}} \, ,
\label{TRC}
\ee
where $\Delta S_{CH} = S_{CH}(M +\omega, Q) -S_{CH}(M, Q)$ is the change of entropy of the CH after and
before the particle tunnels into the CH. Once again, the actual radiant spectrum is no longer precisely
thermal.

\section{Radiation of massive charged particles via tunnelling}

Now, we proceed to discuss Hawking radiation of charged massive particles from the EH. Since the emitted
particle can be treated as a shell of energy $\omega$ and charge $q$, Eqs. (\ref{PLE}) and (\ref{MCRGL})
should be modified when the particle's self-gravitation is incorporated. Taking into account the energy
conservation and the charge conservation, the mass and charge parameters in these equations will be replaced
with $M\to M -\omega$ and $Q\to Q -q$ when the particle of energy $\omega$ and charge $q$ tunnels out of
the EH. So the outgoing radial geodesics of the charged massive particle tunnelling out from the EH and
the non-zero component of electro-magnetic potential are, respectively,
\bea
\dot{r} &=& \frac{f(M -\omega, Q -q, r)}{2\sqrt{1 -f(M -\omega, Q -q, r)}}
\approx \kappa_+(M -\omega, Q -q, r_+)\big(r -r_+\big) \, , \\
A_t &=& \frac{Q -q}{(n-1)V_nr^{n-1}} \, .
\eea

When we investigate the tunnelling process of a charged massive particle, the effect of the electro-magnetic
field outside the black hole should be taken into consideration. So the matter-gravity system consists of
the black hole and the outside electro-magnetic field whose Lagrangian function $-(1/4)F_{\mu\nu}F^{\mu\nu}$
is described by the generalized coordinate $A_{\mu} = (A_t, 0, 0, 0)$. As the generalized coordinate $A_t$
is an ignorable one, to eliminate this degree of freedom completely, the imaginary part of the action should
be written as
\be
\mathrm{Im} S = \mathrm{Im} \int_{t_i}^{t_f}\big(L -P_{A_t}\dot{A_t}\big)dt
 = \mathrm{Im}\int_{r_i}^{r_f}\Big[\int_{(0, ~0)}^{(P_r, P_{A_t})}
\big(\dot{r}~dP_r^{\prime} -\dot{A_t}~dP_{A_t}^{\prime}\big)\Big]\frac{dr}{\dot{r}} \, ,
\label{IA}
\ee
where $r_{i}$ and $r_{f}$ represent the locations of the EH before and after the particle of energy $\omega$
and charge $q$ tunnels out, and ($P_{A_t}, P_r$) are two canonical momenta conjugate to the coordinates
($A_t, r$), respectively.

Substituting the Hamilton's equations of motion
\bea
\dot{r} &=& \frac{dH}{dP_r}\Big|_{(r; A_t, P_{A_t})} \, , \qquad
dH|_{(r; A_t, P_{A_t})} = d\big(M -\omega\big) \, , \\
\dot{A_t} &=& \frac{dH}{dP_{A_t}}\Big|_{(A_t; r, P_r)} \, , \qquad
dH|_{(A_t; r, P_r)} = A_td(Q -q) \, ,
\eea
into Eq. (\ref{IA}), and switching the order of integration yield the imaginary part of the action
\bea
\mathrm{Im}S &=& \mathrm{Im}\int_{r_i}^{r_f} \int_{(M, ~Q)}^{(M -\omega, ~Q -q)}
\Big[d(M -\omega^{\prime}) -\frac{Q -q^{\prime}}{(n-1)V_nr^{n-1}}
d(Q -q^{\prime})\Big] \frac{dr}{\dot{r}} \nn \\
&=& \mathrm{Im}\int_{(M, ~Q)}^{(M -\omega, ~Q -q)}\int_{r_i}^{r_f}
\frac{2\sqrt{1 -f(M -\omega^{\prime}, Q -q^{\prime}, r)}}{f(M -\omega^{\prime}, Q -q^{\prime}, r)}
\Big[d(M -\omega^{\prime}) \nn \\
&&\qquad  -\frac{Q -q^{\prime}}{(n-1)V_nr^{n-1}}d(Q -q^{\prime})\Big]dr \, .
\eea

The above integral can be evaluated by deforming the contour around the single pole $r = r_+^{\prime}$ at
the EH. Doing the $r$ integral first, we find
\bea
\mathrm{Im}S &=& -\int_{(M, ~Q)}^{(M -\omega, ~Q -q)}
\frac{\pi}{\kappa_+(M -\omega^{\prime}, Q -q^{\prime}, r_+^{\prime})}\Big[d(M -\omega^{\prime}) \nn \\
&&\qquad -\frac{Q -q^{\prime}}{(n-1)V_nr_+^{\prime (n-1)}} d(Q -q^{\prime})\Big] \, .
\eea
Using the area-entropy formulae of the EH: $S_{EH} = V_nr_+^n/4 = 4\pi r_+^n/(n\omega_n)$, one can prove
an identity which is essentially equivalent to the differential form of the first law of black hole
thermodynamics
\be
\frac{\pi}{\kappa_+(M, Q, r_+)}\Big[dM -\frac{Q}{(n-1)V_nr_+^{n-1}} dQ\Big]
 = \frac{2\pi}{\omega_n}r_+^{n-1}dr_+ \equiv \frac{1}{2}dS_{EH} \, .
\label{DFL}
\ee
By means of this identity, we can easily finish the integration and obtain
\bea
\mathrm{Im}S &=& -\int_{r_i}^{r_f}\frac{2\pi}{\omega_n}r_+^{\prime (n-1)}dr_+^{\prime}
 = -\frac{2\pi}{n\omega_n}\big(r_f^n -r_i^n\big) \nn \\
&\equiv& -\frac{1}{2}\int_{(M, ~Q)}^{(M -\omega, ~Q -q)}dS^{\prime}
 = -\frac{1}{2}\Delta S_{EH} \, .
\eea
The tunnelling probability for the outgoing radiation at the EH is still represented by Eq. (\ref{TRE}) with
$\Delta S_{EH} = S_{EH}(M -\omega, Q -q) -S_{EH}(M, Q)$ being the difference of black hole entropy after and
before the particle emission. Obviously, the derived emission spectrum actually deviates from pure thermality.

Finally, we turn to calculate the tunnelling rate at the CH. Different from the tunnelling behavior across
the EH, the charged particle is found to tunnel into the CH. Ignoring the effect of the EH and taking into
account the energy conservation and electric charge conservation, the mass and charge parameters in Eqs.
(\ref{PLE}) and (\ref{MCRGL}) will be replaced with $M\to M +\omega$ and $Q\to Q +q$, and generally speaking,
the size of EH will expand and the radius of CH will shrink when the charged particle tunnels into the CH.
At this moment, the radial geodesics of the charged massive particle tunnelling into the CH and the
non-vanishing component of electro-magnetic potential are
\bea
\dot{r} &=& -\frac{f(M +\omega, Q +q, r)}{2\sqrt{1 -f(M +\omega, Q +q, r)}}
\approx -\kappa_c(M +\omega, Q +q, r_c)\big(r -r_c\big) \, , \\
A_t &=& -\frac{Q +q}{(n-1)V_nr^{n-1}} \, .
\eea

It should be stressed that the first law of thermodynamics is satisfied both at the EH and CH, respectively,
but one must note that the definitions of energy, temperature, and electric potential receive an opposite
sign in both cases \cite{CRG}. With this issue in mind, the calculation of the tunnelling rate at the CH
is almost similar to those did in the EH case. In this way, one can get
\bea
\mathrm{Im}S &=& -\mathrm{Im}\int_{r_{ic}}^{r_{fc}} \int_{(-M, -Q)}^{(-M -\omega, -Q -q)}
\Big[d(M +\omega^{\prime}) -\frac{Q +q^{\prime}}{(n-1)V_nr^{n-1}}
d(Q +q^{\prime})\Big] \frac{dr}{\dot{r}} \nn \\
&=& \mathrm{Im}\int_{(-M, -Q)}^{(-M -\omega, -Q -q)} \int_{r_{ic}}^{r_{fc}}
\frac{2\sqrt{1 -f(M +\omega^{\prime}, Q +q^{\prime}, r)}}{f(M +\omega^{\prime}, Q +q^{\prime}, r)}
\Big[d(M +\omega^{\prime}) \nn \\
&&\qquad -\frac{Q +q^{\prime}}{(n-1)V_nr^{n-1}}d(Q +q^{\prime})\Big]dr \, ,
\eea
where $r_{ic}$ and $r_{fc}$ are the locations of the CH before and after the charged particle tunnels into
the CH, respectively. Carrying out the integration, we have
\bea
\mathrm{Im}S &=& -\int_{(-M, -Q)}^{(-M -\omega, -Q -q)}
\frac{\pi}{\kappa_c(M +\omega^{\prime}, Q +q^{\prime}, r_c^{\prime})} \Big[d(M +\omega^{\prime}) \nn \\
&&\qquad -\frac{Q +q^{\prime}}{(n-1)V_nr_c^{\prime (n-1)}}d(Q +q^{\prime})\Big]  \nn \\
 &=& -\frac{1}{2}\int_{(M, ~Q)}^{(M +\omega, ~Q +q)}dS^{\prime}
 = -\frac{1}{2}\Delta S_{CH} = -\frac{2\pi}{n\omega_n}\big(r_{fc}^n -r_{ic}^n\big) \, ,
\eea
in which we have used an identity similar to Eq. (\ref{DFL}) that holds at the CH. Thus the tunnelling rate
is still given by Eq. (\ref{TRC}) where $\Delta S_{CH} = S_{CH}(M +\omega, Q +q) -S_{CH}(M, Q)$ is the change
of entropy of the CH after and before the particle tunnels into the CH. Once again, the actual radiant
spectrum is no longer precisely thermal and fits well to the universal relation presented in the standard
semi-classical tunnelling formalism \cite{PW,KKW}.

\section{Concluding remarks}

In summary, we have investigated Hawking radiation of the massless uncharged particles and massive charged
particles as a semi-classical tunnelling process from the black hole event horizon and the cosmological
horizon of higher dimensional Reissner-Nordstr\"{o}m black holes in de Sitter space-time, respectively.
Our results show that the tunnelling rate is related to the change of Bekenstein-Hawking entropy and
the exact radiant spectrum is no longer precisely thermal after considering the black hole background as
dynamical and incorporating the self-gravitation effect of the emitted particles when the energy conservation
and electric charge conservation are taken into account, thus perfectly extending the Kraus-Parikh-Wilczek's
semi-classical tunnelling framework in the four-dimensional case to the higher dimensional spherically
symmetric case.

The leading-order, energy-independent term in the emission spectrum is the well-known pure thermal spectrum,
but the energy-dependent terms when computed to any desired order in $\omega$, are indicative of a ``greybody''
factor in the emission spectrum. This means that a deviation from pure thermality probably have an important
effect on the number of the brane localized particles emitted per unit time. This issue deserves further
research in the future.

\section*{Acknowledgments}

S.-Q.Wu was supported by a starting fund from Central China Normal University and by the Natural Science
Foundation of China. The authors thank Dr. E.C. Vagenas for pointing out some misprints.

\end{document}